\documentclass{article}

\usepackage{color,graphicx,epsfig}
\begin{document}
ZU-TH 19/98
\vskip5mm
\hrule
\vskip3mm
\begin{center}
\large{{\bf ESA Conference 
\vskip2mm
The Universe as seen by ISO
\vskip2mm
UNESCO-Paris (France), 20-23 October 1998}}
\end{center}
\vskip3mm
\hrule
\vskip2cm
\begin{center}
\huge{\bf Molecules at High Redshifts} 
\vskip8mm
\large{\bf Denis Puy$^{1,2}$ and Monique Signore$^3$}
\end{center}
\vskip2mm
\noindent
\begin{center}
$^1$ Paul Scherrer Institute\\
Laboratory for Astrophysics\\
CH-5232 Villigen-PSI (Switzerland)
\vskip2mm
\noindent
$^2$ Institute of Theoretical Physics, University of Zurich\\
Winterthurerstrasse, 190\\
CH-8050 Zurich (Switzerland)\\
puy@unizh.physik.ch
\vskip2mm
\noindent
$^3$Observatoire de Paris\\
D\'epartement de Radioastronomie Millim\'etrique\\
61, Avenue de l'observatoire\\
75014 Paris (France)\\
signore@mesiob.obspm.fr
\vskip2mm
\end{center}
\clearpage
\section{Introduction}
The study of primordial 
chemistry of molecules adresses a number of interesting questions
pertaining to the thermal balance of collapsing molecular
protoclouds. 
In numerous astrophysical cases molecular cooling and heating influence 
dynamical evolution of the medium
\\
The existence of a
significant abundance of molecules can be crucial on the dynamical
evolution of collapsing objects. Because the cloud temperature 
increases with contraction, a cooling mechanism can be important for
the structure formation, 
by lowering pressure opposing gravity, i.e. by allowing
continued collapse of Jeans unstable protoclouds. This is particularly
true for the first generation of objects. 
\\
It has been suggested that a finite amount of molecules such as $H_2$,
$HD$ and $LiH$ can be formed immediately after the recombination of
cosmological hydrogen (Lepp and Shull 1984, Puy et al 1993). 
The cosmological fractional abundance $HD/H_2$ is small. 
Nevertheless, the presence of a
non-zero permanent electric dipole moment makes $HD$ a
potentially more important coolant than $H_2$ at modest temperatures, 
 although $HD$ is much less abundant than $H_2$. Recently
Puy and Signore (1997) showed that during
the early stages of gravitational collapse, for some collapsing masses, 
$HD$ molecules were the main cooling agent when the collapsing
protostructure had a temperature of about 200 Kelvins.
\\
Thus in section 2,  we analytically estimate the
molecular cooling during the collapse of protoclouds Then, in section
3, the possibility of thermal instability during the early phase of
gravitational collapse is discussed. 
\section{Molecular functions of a collapsing protocloud}
We consider here only the transition between the ground state and the
first state. The transition is a dipolar transition
whereas for $H_2$ molecule the transition is quadrupolar (because it
has no permanent dipolar moment).
\\
In this study, we consider only the molecular cooling which is
the dominant term in the cooling function as we have shown for
a collapsing cloud (Puy \& Signore 1996, 1997, 1998a, 1998b).
 we obtain for the cooling
function of $HD$ the expression:
\begin{equation}
\Lambda_{HD} \, \sim \,
\frac{3 C_{1,0}^{HD} exp(-\frac{T_{1,0}^{HD}}{T_m})
 h \nu_{1,0}^{HD} n_{HD} A_{1,0}^{HD}}
{A_{1,0}^{HD} +
C_{1,0}^{HD} \Bigl [ 1 + 3 exp(-\frac{T_{1,0}^{HD}}{T_m}) \Bigr ]}
\end{equation} 

and the molecular cooling for the $H_2$ molecule is given by:
\begin{equation}
\Lambda_{H_2} \, \sim \, 
\frac{
5 n_{H_2} C_{2,0}^{H_2} A_{2,0}^{H_2} h \nu _{2,0}^{H_2} 
exp(-\frac{T_{2,0}^{H_2}}{T_m})
}
{A_{2,0}^{H_2}+C_{2,0}^{H_2} \Bigl [
1+5 exp(-\frac{T_{2,0}^{H_2}}{T_m}) \Bigr ]
}
\end{equation}

The total cooling function is given by 
\begin{equation}
\Lambda_{Total} \, = \, \Lambda_{HD} + \Lambda_{H_2} \, = \, 
\Lambda_{HD} (1 + \xi_{H_2})
\end{equation}
where 
$$
\xi_{H_2} \, = \, \frac{\Lambda_{H_2}}{\Lambda_{HD}}$$ 
is the ratio
between the $H_2$ and $HD$ cooling functions. Finally we deduce:
\begin{equation}
\Lambda_{HD} \, \sim \, 2.66 \times 10^{-21} n^2_{HD} \, 
exp(-\frac{128.6}{T_m})
\frac{\sqrt{T_m}}{1 + n_{HD} \sqrt{T_m} \Bigl [ 1 + 3 
exp(-\frac{128.6}{T_m}) \Bigr] } 
\end{equation} 
\begin{equation}
\Lambda_{H_2} \, \sim \, 1.23 \times 10^{-20} n^2_{HD} \, exp(-\frac{512}{T_m})
\frac{\sqrt{T_m}}{5 \times 10^{-4} + n_{HD} \sqrt{T_m} \Bigl [ 1 + 5 
exp(-\frac{512}{T_m}) \Bigr] } 
\end{equation} 
and for the ratio:
\begin{equation}
\xi_{H_2} \, \sim  \, 
4.63 \, e^{-\frac{383.4}{T_m}} \, \frac{
1+\sqrt{T_m}n_{HD} [ 1 + 3 exp(-\frac{128.6}{T_m})]}
{5 \times 10^{-4} +\sqrt{T_m} n_{HD} [ 1 + 5
exp(-\frac{512}{T_m})]}
\end{equation}

We study a homologous model of spherical collapse of mass
$M$ similar to the model adopted in Lahav (1986) and Puy \& Signore
(1996) in which we only consider $H_2$ and $HD$ molecules. The aim 
of this paper is to analyse, through our approximations (where the
only two first excited levels are considered), the evolution of the cooling and
 the potentiality of thermal instability. 
\\
First, let us recall the equations governing dynamics of a collapsing
protocloud: 
\begin{equation} 
\frac{dT_m}{dt} \, = \, 
-2 \frac{T}{r}.\frac{dr}{dt} - \frac{2}{3n k}\Lambda_{Total}
\end{equation}
for the evolution of the matter temperature $T_m$,
\begin{equation}
\frac{d^2 r}{dt^2} \, = \, \frac{5k T_m}{2 m_H r} - 
\frac{G M}{r^2}
\end{equation}
for the evolution of the radius $r$ of the collapsing cloud and
\begin{equation}
\frac{dn}{dt} \, = \, -3 \frac{n}{r}.\frac{dr}{dt}
\end{equation}
for the evolution of the matter density $n$.
\\
At the beginning of the gravitational
collapse the matter temperature increases, then due to the very
important efficiency of the molecular cooling, the temperature
decreases. We consider here this {\it transition regime} i.e. the point where
the temperature curve has a horizontal asymptot (see Puy \& Signore 
1997): 
$$
\frac{dT_m}{dt} \, = \, 0
$$

Finally we obtain for the total molecular cooling
\begin{equation}
\Lambda _{Total} \, = \, \delta \, T_m^{7/2} \, 
\frac{exp(5T_o/T_m)}{\Bigl[ 1 + \xi_o exp(-\Delta T_o / T_m)
\Bigr ]^5}
\end{equation}

$\delta = \frac{3 k G M^2 }{ 2 m_H \kappa^5}\, M^2 $ with $\kappa \, = \, 
1.4 \times 10^{22}$, $ T_o= 128.6$ K, $\Delta T_o = 383.4$ K and 
$\xi_o = 9260$. 
\section{Thermal Instability}
We have learned much about thermal instability in the last 30 years since
the appearance of the work of Field (1965). Many studies have
focused on the problem of thermal instabilitie in different situations. In 
general, astronomical objects are formed by self-gravitation. However, some 
objects can not explained by this process. For these objects the gravitational 
energy is smaller than the internal energy. Thus if the thermal equilibrium of 
the medium is a balance between energy gains and radiative losses, instability
results if, near equilibrium, the losses increase with decreasing temprature. 
Then, a cooler than average region cools more effectively than its
surroundings, and its temperature rapidly drops below the initial equilibrium
value.
\\
Thus, if we introduce a perturbation of density and temperature such that some
thermodynamics variable (pressure, temperature...) is held constant, the
entropy of the material $\cal{S}$ will change by an amount $\delta {\cal{S}}$,
and the heat-loss function, by an amount $\delta  \Psi=\Gamma - \Lambda$. From
the equation  $$ \delta \Psi \, = \, -T d(\delta {\cal{S}} )
$$
there is instability only if:
$$
\Bigl ( \frac{\delta \Psi}{\delta {\cal{S}}} \Bigr ) \, > \, 0
$$
We know that $T d{\cal{S}} = C_p dT$ in an 
isobaric perturbation. The corresponding inequality can be written: 
$$
\Big ( \frac{\delta \Psi}{\delta T} \Big )_p 
\, = \, 
\Big ( \frac{\delta \Psi}{\delta T} \Big )_n
+
\frac{\partial n}{\partial T} \,
\Big ( \frac{\delta \Psi}{\delta n} \Big )_T
\, < \, 0
$$
which characterizes the Field's criterion. The criterion involves
constant $P$ because small blobs tend to maintain pressure equilibrium
with their surroundings when heating and cooling get out of balance. 
In our case (i.e. at the transition regime), a thermal
instability could spontaneously be developped if the
Field's Criterion is verified:
\begin{equation}
\frac{\partial \, ln \Lambda }{\partial \, ln T_m} \, < \, 0
\end{equation} 
In our case the logaritmic differential of the total molecular cooling
function is given by 
\begin{equation}
\frac{\partial \, ln \Lambda}{\partial \, ln T_m} \, = \, 
\frac{
7T_m +7T_m \xi_o e^{-\frac{\Delta T_o}{T_m}}-10 T_o -
10T_o \xi_o e^{-\frac{\Delta T_o}{  T_m}} - 10 \xi_o \Delta T_o 
e^{-\frac{\Delta T_o }{ T_m}}
}
{2 + 2 \xi_o e^{-\frac{\Delta T_o}{ T_m}}}
\end{equation}
In figure 1, we have plotted the curve $y(T_m) \, = \, \partial \, ln
\Lambda / \partial \, ln T_m $. It shows that the {\it
Field's criterion} is always verified. 
\\
We conclude that a thermal instability is possible at the {\it 
transition regime}. Thus by maintening
the same pressure in its surroundings, such a blob would get cooler
and cooler and denser and denser until it could separate into
miniblobs. This possibility is very interesting and could give a
scenario of formation of primordial clouds. However, a quantitative
study is necessary to evaluate the order of magnitude of the
mass and the size of the clouds. This last point is crucial.
\\
Therefore, from these estimations, the calculations of molecular
functions could, in principle, be extended to the $CO$ molecules for
collapsing protoclouds at $z<5$ (see Puy \& Signore 1998b). 
But, in any case -and in particular for the discussion of
thermal instability at the {\it transition regime}- the corresponding
total cooling function must be numerically calculated because the
excitation of many rotational levels must be taken into account. 
This extended study is beyond the scope of our paper. 
\section*{Acknowledgements}
The authors gratefully acknowledge St\'ephane Charlot, Philippe Jetzer, Lukas
Grenacher and Francesco Melchiorri  for valuable discussions on this field. 
Part of the work of D. Puy has been conducted under the auspices 
of the {\it D$^r$ Tomalla Foundation} and Swiss National Science Foundation
\section*{References}
{\footnotesize
\noindent
Field G. B. 1965 ApJ 142, 531
\\
Lahav O. 1986 MNRAS 220, 259
\\
Lepp S., Shull M., 1984, ApJ 280, 465
\\
Puy D., Alecian G., Lebourlot J., L\'eorat J.,
Pineau des Forets G. 1993 A\&A 267, 337 
\\
Puy D., Signore M., 1996, A\&A 305, 371
\\
Puy D., Signore M., 1997, New Astron. 2, 299
\\
Puy D., Signore M., 1998a, New Astron. 3, 27
\\
Puy D., Signore M., 1998b, New Astron. 3, 247
}
\begin{figure}
\begin{center}
\includegraphics[height=0.33\textheight, angle=-90]{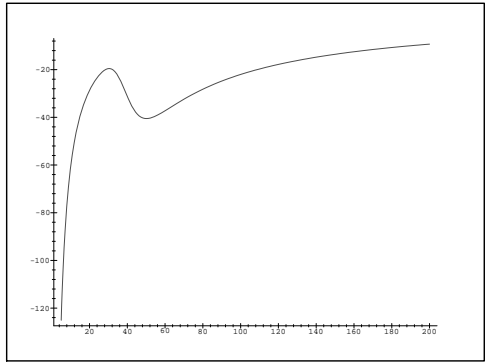}
\caption{\small{$\partial ln \Lambda_{Total}/
\partial ln T_m$ in y-axis and temperature of the matter in x-axis
(in Kelvins) }}
\end{center}
\end{figure}

\end{document}